\newcommand{\be}{\begin{equation}}
\newcommand{\ee}{\end{equation}}
\newcommand{\bea}{\begin{eqnarray}}
\newcommand{\eea}{\end{eqnarray}}
\newcommand{\ba}[1]{\begin{array}{#1}}
\newcommand{\ea}{\end{array}}
\documentclass[twocolumn,secnumarabic,amssymb,nobibnotes,nofootinbib,aps,pra,showpacs]{revtex4}
\usepackage{epsfig}
\usepackage{amssymb}
\begin{document}
\setlength{\topmargin}{-0.05in}

\title{Rotational excitations in two-color photoassociation}
\author{Jisha Hazra$^1$ and Bimalendu Deb$^{1,2}$}
\affiliation{$^1$Department of Materials Science, and 
$^2$Raman Center for Atomic, Molecular and Optical Sciences, Indian Association
for the Cultivation of Science,
Jadavpur, Kolkata 700032, India.}


\def\zbf#1{{\bf {#1}}}
\def\bfm#1{\mbox{\boldmath $#1$}}
\def\hf{\frac{1}{2}}
\begin{abstract}
We show that it is possible to excite higher rotational states $J > 2$ in ultracold photoassociation by two laser fields. Usually higher $J$ states are suppressed in photoassociation at ultracold temperatures in the regime of  Wigner threshold laws. We propose a scheme in which one strong laser field drives photoassociation transition close to either  $J = 1$ or $J = 2$ rotational state of a particular vibrational level of an electronically excited molecule. The other laser field is tuned near photoassociation resonance with $J > 2$ rotational levels of the same vibrational state.  The strong laser field induces a strong continuum-bound dipole coupling. The resulting dipole force between  two colliding atoms modifies the continuum states forming continuum-bound dressed states with a significant component of higher partial waves in the continuum configuration. When the second laser is scanned near the resonance of the higher $J$ states, these states become populated due to photoassociative transitions from the modified continuum.

\end{abstract}

\pacs{34.50.Cx, 34.50.Rk, 42.65.Dr, 33.20.Sn}
\maketitle

\section{introduction}

Photoassociation (PA) spectroscopy \cite{thorsheimPRL5887,julienneRMP7806} of ultracold atoms by which two colliding atoms absorb a photon to form an excited molecular state is an important tool for studying ultracold collisional properties at the interface of atomic and molecular states. PA is particularly useful for producing translationally cold molecules \cite{TsaiPRL7997,PilletPRL8098,Takekoshi,AraujoJCP11903,DeiglmayrPRL10108,LangPRL10108,jingPRA8009} and generating optical Feshbach resonance \cite{FedichevPRL7796,fatemiPRL2002,theis,EnomotoPRL101,debPRL10409}. More than a decade ago, theoretical models \cite{theory1,theory2} were developed to explain PA line shape in the weak-coupling regime. The effects of laser intensity on PA spectra \cite{BohnPRA5697,juliennePRA6099,williamsPRA662002,ZimmermannPRA6602,huletPRL9103,juliennePRA6904} have been an important current issue. Over the years, two-color Raman type PA has emerged as an important method for creating translationally cold molecules in the ground electronic configuration. Recently, using this method, cold polar molecules \cite{DeiglmayrPRL10108} in rovibrational ground state have been produced. Molecules created by one- or two- color PA of ultracold atoms generally possess low-lying rotational levels $J \leq 3$. Motivated by recent experimental observation of excitation of higher rotational states in ultracold PA with an intense laser field \cite{GomezPRA7507}, we here expore theoretically the possibility of rotational excitations in two-color PA. This may be important for producing translationally cold molecules in selective higher rotational states. Previously, two-color PA has been investigated in different other contexts \cite{bagnatoPRL7093,LeonhardtPRA5295,MolenaarPRL7796,JonesJPB3097,MarcassaPRL7394,SuominenPRA5195,ZilioPRL7696,AbrahamPRL7495}, such as photo-ionization of excited molecules \cite{bagnatoPRL7093,LeonhardtPRA5295,MolenaarPRL7796,JonesJPB3097}, shielding of atomic collision \cite{MarcassaPRL7394,SuominenPRA5195,ZilioPRL7696}, measurement of s-wave scattering length \cite{AbrahamPRL7495}, etc.

In this paper we propose a method of two-color photoassociation of two homonuclear atoms for exciting higher rotational levels.  Our proposed method is schematically shown in Fig.1. Laser L$_A$ is a strong field and the laser L$_B$ is a weak one. L$_A$ is tuned near either $J_{A} = 1$ or $J_{A} = 2$ rotational state of a particular vibrational level $v$ of the excited state. This rotational state is predominantly accessed by PA transition from s-wave scattering state. A photon from L$_A$ causes PA excitation from the continnum (s-wave) to the bound level $J_{A}$. A second photon from the same laser can cause a stimulated de-excitation back to the continnum state. This is a stimulated Raman-type process which can lead to significant excitation of higher partial waves in the two-atom continnum. Now, if a weak laser L$_B$ is tuned near $J_{B} > 2$ states, these higher rotational states get excited due to PA from the modified continnum. In this scheme of two-color PA, three photons are involved. This does not fit into a standard $\Lambda$ or V-type process. Here bound-bound transition is absent. All the transitions are of continnum-bound type. This scheme may be viewed as a combination of $\Lambda$ and V-type process with continnum acting as an intermediate state for V-type transition.
\begin{figure}
\includegraphics[angle=270,width=5.00in]{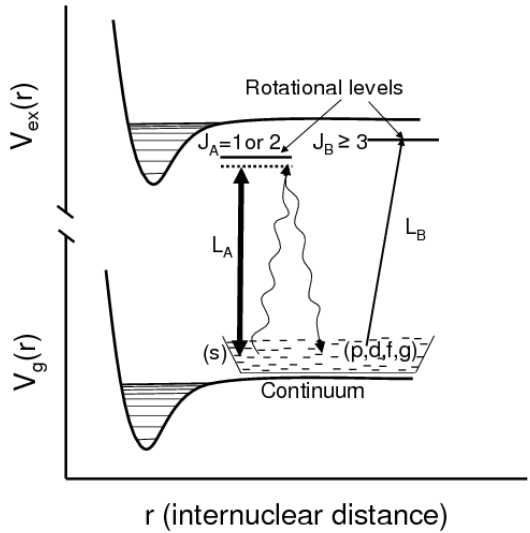}
\caption{A schematic diagram showing the strong (double-arrow thick line) and weak (single-arrow thin line) field couplings between the excited rotational levels $J_{\alpha}$ ($\alpha = A, B$) and the continnum state. Strong laser L$_A$ modifies the continnum state by a two-photon process (curly lines) as described in the text. The laser L$_B$ is tuned near resonance with the rotational levels $J_B \geq $ 3 which are then populated due to PA transition from the modified continnum. Molecular rotational levels $J=1$ and $J=2$ are accessible from s-wave ($\ell = 0$) scattering state, but $J \geq 3$ can only be accessed from higher partial-wave ($\ell > 0$) scattering states.}
\end{figure} 
In the previous Raman-type PA  experiments, excited molecular state is used as an intermediate state. Furthermore, usually two-color PA is carried out in the weak coupling regime. In contrast, our proposed scheme involves necessarily one strong laser field for inducing strong PA coupling. We demonstrate excitation of higher rotational levels in two-color ultracold PA by resorting to a simplified model. We first evaluate higher partial wave scattering states modified due to strong photoassociative coupling \cite{debPRL10409} induced by the strong laser L$_A$. We employ these modified wave functions to calculate two-color stimulated line widths which are significantly enhanced compared to those in the case of one-color.

The paper is organized as follows. In the following section we describe the formulation of the problem and its solution. The numerical results and discussion has been given in Sec.3. Finally the paper is concluded in Sec.4.


\section{The Model and Its Solution}

To start with, let us consider that PA laser couples continuum (scattering) states of collision energy $E = \hbar^2 k^2 /(2\mu)$ (where $\mu$ is the reduced mass) of two alkali-type homo-nuclear ground state $S$ atoms to an excited diatomic (molecular) bound state which asymptotically corresponds to one ground $S$ and another excited $P$ atom. Under electric dipole approximation, the interaction Hamiltonian can be expressed as
\be H_{int} = \sum_{i=1,2} E_{L} \hat{\pi}.\hat{d}_i \label{hintatom} \ee
where $\hat{d}_i = - {\mathrm e} {\mathbf r}_i$ is the dipole moment of $i$-th atom  whose  valence electron's position is given by ${\mathbf r}_i$ with respect to the center of mass of this atom.  Here ${\mathrm e}$ represents an electron's charge, $E_L$ is the laser field amplitude and $\hat{\pi}$ is the polarization vector of the laser. The total Hamiltonian in the center-of-mass frame of the two atoms can be written as
\be
H = H_{elec}({\mathbf r_1},{\mathbf r_2};{\mathbf r_a},{\mathbf r_b})  - \frac{\hbar^2}{2\mu}\nabla_r^2 - \frac{\hbar^2}{2 M}\nabla_R^2 + H_{hf} + H_{int}
\label{hamil}
\ee
where $H_{elec}$ is the electronic part of the Hamiltonian which includes terms corresponding to kinetic energy of the two valence electrons, mutual Coulomb interactions between nuclei and the electrons, exchange  and  electronic spin-orbit interaction. Here ${\mathbf r_a}$ and ${\mathbf r_b}$ represent the position vectors of the nuclei of atoms a and b, respectively; $\nabla_r$ and $\nabla_R$ denote the Laplacian operators corresponding to the relative coordinate
${\mathbf r} = {\mathbf r_a} - {\mathbf r_b}$ and the center-of-mass coordinate ${\mathbf R} = ({\mathbf r_a} +{\mathbf r_b})/2$ and $H_{hf}$ stands for the hyperfine interaction of two atoms. Under Born-Oppenheimer approximation, while solving the electronic part of the Hamiltonian, the nuclear coordinates appear merely as  parameters. PA laser couples only two electronic molecular states which are the initial ground and the final excited diatomic states  represented by $\langle {\mathbf r}_1,{\mathbf r}_2;r \mid g \rangle = \phi_g({\mathbf r}_1,{\mathbf r}_2;  r) $ and $\langle {\mathbf r}_1,{\mathbf r}_2;r\mid e \rangle = \phi_e({\mathbf r}_1,{\mathbf r}_2; r) $, respectively. These internal electronic states have parametrical dependence on the internuclear coordinate $r$. They satisfy the eigenvalue equations
\be
H_{elec} \phi_{\alpha} ({\mathbf r_1},{\mathbf r_2};  r) = V_{\alpha} (r) \phi_{\alpha} ({\mathbf r_1},{\mathbf r_2}; r); \hspace{1cm} \alpha = g,e
\ee
We assume that the matrix element $ \langle e \mid H_{int} \mid g \rangle \simeq \Omega_{eg}(r)$ depends only on separation $r$. Then the center-of-mass motion gets decoupled from relative motion. Henceforth we consider only the relative motion. By specifying the electronic parts of both the bound and the  continuum states, one can calculate the matrix element of $H_{int}$ over the electronic parts of the two molecular levels involved in free-bound transition and thus obtain molecular coupling strength $\Lambda(r)$.

The continnum-bound dressed state can be written as $\Psi_{E}({\mathbf r_1},{\mathbf r_2}; {\mathbf r}) = \sum_{\alpha = g,e} \Phi_{\alpha}({\mathbf r}) \mid \alpha \rangle$ which is assumed to be energy normalized with $E$ being the energy eigen value. In the absence of atom-field interaction $H_{int}$, the problem is to find out multi-channel scattering wave-function in the ground electronic configuration. The scattering channels correspond  to two separated atoms a and b in hyperfine spin $f_a$ and $f_b$, respectively. The molecular hyperfine state is characterized by the spin $\overrightarrow F = \overrightarrow f_a + \overrightarrow f_b$. A channel is defined by the angular state $\mid {\cal F}; f_a, f_b, \ell \rangle$ where $\overrightarrow{\cal F} = \overrightarrow F + \overrightarrow  \ell = \overrightarrow f_a + \overrightarrow f_b + \overrightarrow \ell$ where $\ell$ is the mechanical angular momentum of the relative motion of the two atoms. This asymptotic basis $\mid {\cal F}; f_a, f_b, \ell \rangle$ can be expressed in terms of the adiabatic molecular basis $\mid {\cal F}; S, I, \ell \rangle$ \cite{Tiesinga}, where $S$ and $I$ are the total electronic and nuclear spin angular momentum of two atoms. In the case of excited molecular state, $S$ should be replaced by electronic angular momentum $ J_{e} = S + L$. Alternatively, the adiabatic basis $\mid {\cal F}; J_{e}, I, \ell \rangle$ can also be expressed in terms of $\mid {\cal F}; J (J_{e},\ell) I\rangle$. Thus the rotational state of a diatom can be expressed  in terms of the matrix element $ \mid J \Omega M \rangle = i^J \sqrt{\frac{2 J + 1}{8\pi^2}} {\cal D}^{(J)}_{M \Omega }(\hat{r})$ where $M$ and $\Omega$ are the z-component of $J$ in the  space-fixed and body-fixed coordinate frame and $\hat{r}$ represents the Euler angles for transformation from body-fixed to space-fixed frame. ${\cal D}^{(J)}_{M \Omega }(\hat{r})$ is the rotational matrix element. For ground electronic configuration, we have $J = \ell, M = m_{\ell}$ and $\Omega = 0$; thereby, ${\cal D}^{(J)}_{M \Omega }(\hat{r})$ reduces to spherical harmonics $Y_{\ell {m}_{\ell}}$. We thus express the ground state $\Phi_g({\mathbf r})$ in the following form

\be \Phi_g({\mathbf r}) \propto   r^{-1} \sum_{\ell, m_{\ell}}  \left [ \int_{E'} \beta_{E'} \psi_{E' \ell m_{\ell}}( r )d E' \mid \ell m_{\ell} 0 \rangle \right ]  \label{grstate}\ee

where $\psi_{E'\ell m_{\ell}}(r)$ is the energy-normalized scattering state with  collision energy $E'$ and  $\beta_E'$ is the density of states of unperturbed continnum. Similarly, for a particular value of $\Omega$, we can expand the excited state $\Phi_e({\mathbf r}) $ in the following form
\be \Phi_e({\mathbf r}) \propto   r^{-1} \sum_{M} \left [\phi_{vJ}(r) \mid J \Omega M  \rangle \right ] \label{exstate} \ee
Substitution of Eqs. (\ref{grstate}) and (\ref{exstate}) into time-independent Schr\"{o}dinger equation leads to coupled differential equations. These equations are solved by the use of real space Green's function. The detailed method of solution for a model problem is given in Appendix A. In our model calculations, we consider only a single ground hyperfine channel. The solution $\phi_{vJ}(r)$ can be expressed as
\be
\phi_{vJ}(r) = \int_{E'} \beta_{E'}\sum_{\ell m_{\ell} M} A_{JM; \ell m_{\ell}}\phi_{vJ}^{0}(r) dE'
\label{exsol}\ee
where $\phi_{vJ}^{0}(r)$ is the excited molecular state (unit-normalized) in the absence of laser field and \bea A_{J,M; \ell, m_{\ell}} &=&  \left [ f_{J, M; \ell m_{\ell}} + E_{J \ell}^{shift}  \tilde{A}_J \right ] \nonumber \\ &\times& \frac{1}{\hbar\delta+E-E_{vJ} + i \hbar \gamma/2}\label{A_JM:lm_l} \eea is the probability amplitude of excitation of $J$ from a particular partial wave $\ell$. Here
\be f_{J M; \ell m_{\ell}} =  \int\phi_{vJ}^{0}(r') \Lambda_{J M; \ell m_{\ell}}(r') \psi_{E\ell}^{0,reg}(r') d r'\label{F_JM:lm_l}\ee is the continnum-bound dipole matrix element and $\Lambda_{J M; \ell m_{\ell}} =  \langle \ J M \Omega  \mid \Omega_{eg} \mid \ell m_{\ell} 0  \rangle$. $\psi_{E\ell}^{0,reg}(r')$ represents the $\ell$-th partial wave regular scattering solution in the absence of laser field and
\bea E_{J \ell}^{{\mathrm shift}} &=&  \pi \int\int d r' d r \phi_{vJ}^{0}(r') \Lambda_{J M; \ell m_{\ell}}(r')\nonumber \\ &\times& [{\cal K_{\ell}}(r',r)] \Lambda_{\ell m_{\ell}; J M}(r)\phi_{vJ}^{0}(r) \label{Shift_Jl}\eea is the partial light shift of the excited state. Here ${\cal K_{\ell}}(r,r')$ is the propagator as defined in the Appendix A. The total probability amplitude of excitation  $\tilde{A}_J$ for a particular $J$ is given by \be \tilde{A}_J = \sum_{\ell, m_{\ell}, M} \frac{
f_{J M; \ell m_{\ell}}}{\hbar\delta+E-E_{vJ} + i \hbar \gamma/2 -
E_{J}^{shift}} \ee  where $E_J^{shift} = \sum_{\ell, m_{\ell}, M} E_{J \ell}^{shift}$ is the total energy shift of the excited level. $\hbar \gamma/2$  is the natural line width of the excited molecular state,  $E_{vJ}$ is the bound state energy corresponding to the bound state solution $\phi_{vJ}^{0}$ of the excited state. $ \delta = \omega_{L} - \omega_{A} $ is the frequency off-set  between  the laser frequency $\omega_L$ and atomic resonance frequency $\omega_{A}$. The ground state scattering solution in the presence of PA laser is given by 
\bea
\psi_{E \ell m_{\ell}}( r ) &=& \psi_{E\ell}^{0,reg} + \sum_{\ell' m_{\ell'} M } A_{ J M; \ell' m_{\ell'}}(E) \nonumber \\ &\times& \int {\cal K_{\ell}}(r,r')\Lambda_{ \ell m_{\ell}; J
M}(r')\phi_{vJ}^{0}(r')dr' \label{nscf1}\eea In the asymptotic limit ($r\rightarrow \infty$), the modified scattering wave-function behaves like
\bea
\psi_{E \ell m_{\ell}} &=& \cos\eta_{\ell}^{L} \psi_{E\ell}^{0,reg} +  \sin\eta_{\ell}^{L} \psi_{E\ell}^{0,irr}  \label{modiwave}\eea where $\psi_{E\ell}^{0,irr}$ is the irregular wave function of $\ell$-th partial wave.
\begin{table}
\caption{Numerically calculated rotational energies $E_{vJ}$ (in
unit of GHz) and total shift $E_J^{{\mathrm shift}}$ (in
unit of MHz) for one-color laser intensity $I$ = 1
kW/cm$^2$ for vibrational state $v= 48$ of 1$_g$ excited state. 
Also given are the rotational energy spacings $\Delta_{J} = E_{v J} - E_{v J - 1}$ (in
unit of GHz) for a few lowest $J$ values.}
\begin{tabular}{c  c  c  c  c  c  c  c  c  c  c  c  c}
\hline
\multicolumn{1}{c}{$J$ $\longrightarrow$} & \multicolumn{1}{c}{\vline} &  \multicolumn{1}{c}{1} & \multicolumn{1}{c} {\vline} &  \multicolumn{1}{c}{2} & \multicolumn{1}{c} {\vline} & \multicolumn{1}{c}{3} & \multicolumn{1}{c} {\vline} & \multicolumn{1}{c}{4} & \multicolumn{1}{c} {\vline} & \multicolumn{1}{c}{5} & \multicolumn{1}{c} {\vline} & \multicolumn{1}{c}{6} \\
\hline
$E_{vJ}$ (GHz) & \vline  & 0.57  & \vline  & 2.13 & \vline  & 4.77 & \vline  & 8.55 & \vline  & 13.03 & \vline  & 18.34\\
$\Delta_{J}$ (GHz) & \vline  & --  & \vline  & 1.56 & \vline  & 2.64 & \vline  & 3.78 & \vline  & 4.48 & \vline  & 5.31\\
$-E_J^{{\mathrm shift}}$ (MHz) & \vline  & 19.69  & \vline  & 22.79 & \vline  & 17.36 & \vline  & 11.61 & \vline  & 6.83 & \vline  & 2.22\\
\hline
  \end{tabular}
\label{tb1}
\end{table}
Here $\eta_{\ell}^{L}$ is the phase shift due to the applied laser field and is given by
\bea
\tan\eta_{\ell}^{L} &=& -\pi \sum_{\ell' m_{\ell'} M } A_{ J M; \ell' m_{\ell'}}(E) f_{\ell m_{\ell}; J M}\\ \nonumber &=& -\pi \sum_{\ell', m_{\ell'}, M} \frac{
f_{J M; \ell' m_{\ell'}}}{\hbar\delta+E-E_{vJ} + i \hbar \gamma/2 -
E_{J}^{shift}}\nonumber \\ &\times& f_{\ell m_{\ell}; J M}\label{newphase}\eea
where $f_{\ell m_{\ell}; J M} = \int_{0}^{\infty} \phi_{vJ}^{0}(r') \Lambda_{\ell m_{\ell} ; J M }(r') \psi_{E\ell}^{0,reg}(r') d r'$. The two-color partial stimulated line width $\Gamma^{(2)}_{J_B\ell}$ for a particular rotational state $J_B$ is given by \bea \Gamma^{(2)}_{J_B\ell} = 2\pi \left |\int \phi_{vJ}^0 (r)
\Lambda_{J M; \ell m_{\ell}} (r) \psi_{E \ell m_{\ell}}( r ) d r \right|^2 \label{gama_Jl^2}\eea and the total stimulated line width is $\Gamma^{(2)}_{J_B}$ = $\sum_{\ell m_{\ell} M} \Gamma^{(2)}_{J_B\ell}$. The excitation of particular rotational state $J$ from the partial wave $\ell$ is governed by the following selection rule \bea \left | J -|\overrightarrow L+ \overrightarrow S| \right| \le \ell \le \left | J +|\overrightarrow L+ \overrightarrow S| \right| \label{selectionrule}\eea where $L$ is the total electronic orbital angular momentum $\overrightarrow L = \overrightarrow l_{1}+\overrightarrow l_{2}$ and $S$ is the sum of two individual atomic spin, i.e. $\overrightarrow S = \overrightarrow s_{1}+\overrightarrow s_{2}$.
So the lowest possible partial wave $\ell$ which can make the largest contribution to the excitation of rotational state $J$ = 1, 2, 3, 4, 5, 6  are 0, 0, 1, 2, 3, 4, respectively. 
The two-color photoassociation rate $K_{PA}^{(2)}$ for $J_B>2$ is defined as 
\bea
K_{PA}^{(2)}=\langle v_{rel}\sigma_J\rangle=\frac{1}{hQ_T}\int_{0}^{\infty}\hbar P_{J_B}^{(2)} e^{-\beta E}  dE
\label{Photorate}\eea
where $P_{J_B}^{(2)}$ = $\gamma \Gamma^{(2)}_{J}/[(\hbar \delta_{B}+E-E_{v,J_B}-E_{J_B}^{shift})^2+(\gamma+\Gamma^{(2)}_{J})^{2}/4]$ and $v_{rel} = \hbar k/\mu$ is the relative velocity of two atoms, $\sigma_J = \hbar P_{J_B}^{(2)}/k^2$ is the inelastic cross-section due to loss of atoms. Here $\langle \cdots \rangle$ implies an averaging over the distribution of initial velocities, $Q_T = (2\pi \mu K_B T/h^2)^{3/2}$ is the translational partition function and $\beta= (K_B T)^{-1}$. In the next section, we apply this formalism to a model system and obtain numerical results.
\section{Results and Discussion}
\begin{figure}
\includegraphics[width=2.50in]{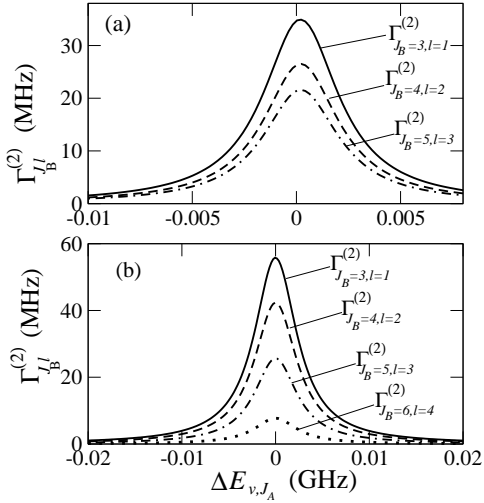}
\caption{Two-color partial stimulated line widths $\Gamma^{(2)}_{J_B\ell}$ (in unit of MHz) as a function of $\Delta E_{v,J_{A}}$  (in unit of GHz) at collisional energy E = 10 $\mu$K. The intensity $I_A$ of laser L$_A$ tuned near $J_A=1$ (a) and $J_A=2$ (b) is 40 kW/cm$^2$ and the intensity $I_B$ of weak laser L$_B$ is 1 W/cm$^2$. The total shift $E_{J_A}^{shift}$ of the rotational state $J_A=1$ and $J_A=2$ are -0.79 GHz and -0.91 GHz, respectively.}
\end{figure} 
\begin{table}
\caption{Tabulated are one- and two-color partial stimulated line widths $\Gamma^{(0)}_{J\ell}$ and $\Gamma^{(2)}_{J_B\ell}$  at $E$ = 10 $\mu$K for two $\delta_A$ values. Here laser $L_A$ is tuned near  $J_A = 1$ rotational state. The intensities of the two lasers are $I_B = 1$ W/cm$^2$ and $I_{A}$ = 40 kW/cm$^2$.}
\begin{tabular}{c c c c c c c c c}
\hline
\multicolumn{1}{c}{} & \multicolumn{1}{c}{} & \multicolumn{1}{c}{} & \multicolumn{1}{c}{} & \multicolumn{1}{c}{} &
\multicolumn{1}{c}{\vline}  & \multicolumn{1}{c}{$\delta_{A}$ = -1.25 GHz}  & 
 \multicolumn{1}{c}{\vline} &\multicolumn{1}{c}{$\delta_{A}$ = -1.48 GHz}\\
\hline
\multicolumn{1}{c}{ $J$ } & \multicolumn{1}{c}{\vline} & \multicolumn{1}{c}{$\ell$} &\multicolumn{1}{c}{\vline}
& \multicolumn{1}{c}{ $\Gamma^{(0)}_{J\ell}$ (MHz) }
& \multicolumn{1}{c}{\vline} & \multicolumn{1}{c}{ $\Gamma^{(2)}_{J_B\ell}$ (MHz) } & \multicolumn{1}{c}{\vline}
& \multicolumn{1}{c}{ $\Gamma^{(2)}_{J_B\ell}$ (MHz) }  \\
\hline
3 & \vline & 1 & \vline & 1.55$\times 10^{-05}$ & \vline & 0.0158 & \vline & 0.0107\\
3 & \vline & 2 & \vline & 0.0000 & \vline & 0.0111 & \vline & 0.0065\\
3 & \vline & 3 & \vline & 0.0000 & \vline & 0.0085 & \vline & 0.0051\\
4 & \vline & 2 & \vline & 0.0000 & \vline & 0.0128 & \vline & 0.0075\\
4 & \vline & 3 & \vline & 0.0000 & \vline & 0.0072 & \vline & 0.0047\\
5 & \vline & 3 & \vline & 0.0000 & \vline & 0.0103 & \vline & 0.0061\\
\hline
  \end{tabular}
\label{tb3}
\end{table}
\begin{figure}
\includegraphics[width=3.41in]{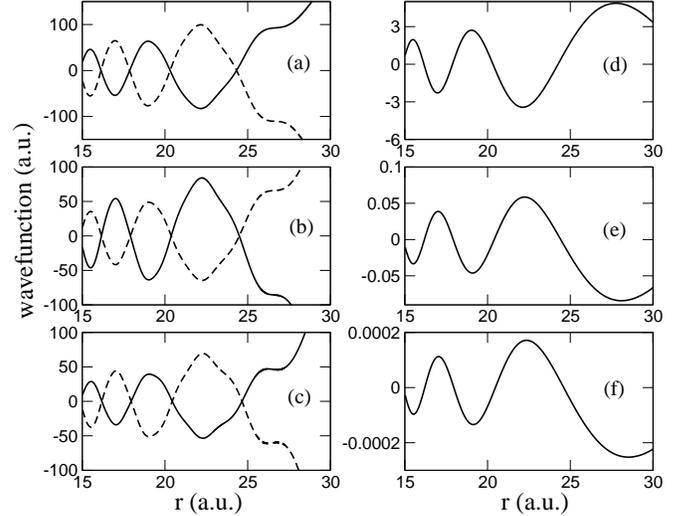}
\caption{The subplots (a), (b) and (c) show the light-induced scattering wavefunctions  $\psi_{E \ell m_{\ell}}$ (in unit of Bohr radius$^{-1/2}$ $\times$ Hartree$^{-1/2}$) for p- ($\ell = 1 $), d- ($\ell = 2$) and f-wave ($\ell = 3$), respectively. The solid and dashed curves correspond to the detunning $\delta_{A} = -1.25$ GHz   and $ \delta_{A} = -1.48$ GHz, respectively. The subplots (d), (e) and (f) exhibit the corresponding field-free regular wavefucntions $\psi^{0,reg}_{E\ell}$. All the wave functions are plotted at collisional energy of 10 $\mu$K and intensity I$_A =$ 40 kW/cm$^2$.}
\end{figure}

For numerical illustration, we consider a model system of two cold ground state ($S_{1/2}$) $^{23}$Na atoms undergoing PA transition from ground state $^{3}\Sigma^{+}_{u}$ to the vibrational state $v = 48$ of the excited molecular $1_g$ state \cite{GomezPRA7507}. At large internuclear distance this $1_g$ potential correlates to $^{2} S_{1/2}$ + $^{2} P_{3/2}$ free atoms and at short range to  1 $^{1}\Pi_{g}$ Born-Oppenheimer potential. In Ref. \cite{GomezPRA7507} higher rotational lines upto $J=6$ have been clearly observed in PA with an intense laser field. The centrifugal barrier of $\ell > 0$ of the two-atoms lies at $r > 50a_0$ ($a_0$= Bohr radius) whereas PA excitations occur at $r \sim 27a_0$. Therefore, the higher rotational states will be unlikely to be populated by PA transitions from $\ell >0$ partial-wave scattering states at ultra-cold temperatures in the weak-coupling regime. Previously, higher rotational levels have been excited in PA spectroscopy due to resonant dipole-dipole interaction with transition occurring at large internuclear separations \cite{longrangerotexcitation,longrangeforce}. The numerically calculated rotational energies $E_{vJ}$, energy shifts $E_{J}^{shift}$  and the corresponding energy difference $\Delta_{J} = E_{v J} - E_{v J - 1}$ for six lowest $J$ values are given in Table I. To demonstrate the working of our proposed scheme, we resort to a simplified two-state calculation. We consider only one ground hyperfine channel with $F=4, f_a=2$ and $f_b=2 $ in the absence of any external magnetic field. In the excited molecular state, we neglect the hyperfine interaction. The two-color partial stimulated line width $\Gamma^{(2)}_{J_B\ell}$ is plotted as a function of detunning $\Delta E_{v,J_{A}} = \hbar\delta_A+E-E_{v,J_A}-E_{J_A}^{shift}$ in Fig.2 for $J_B$ ranging from 3 to 6. The strong laser L$_A$ is tuned near $J_A = 1$ (Fig.2a) and $J_A = 2$ (Fig.2b). From Fig.2 we notice that $\Gamma^{(2)}_{J_B\ell}$ strongly depends on the detunning $\Delta E_{v,J_{A}}$ of the strong laser from PA resonance of the rotational level $J_A$. The maximun of $\Gamma^{(2)}_{J_B\ell}$ occurs at $\Delta E_{v,J_{A}}$ = 0. For lower $J_B$ values, the probability of rotational excitation is higher.
\begin{figure}
\includegraphics[width=3.41in]{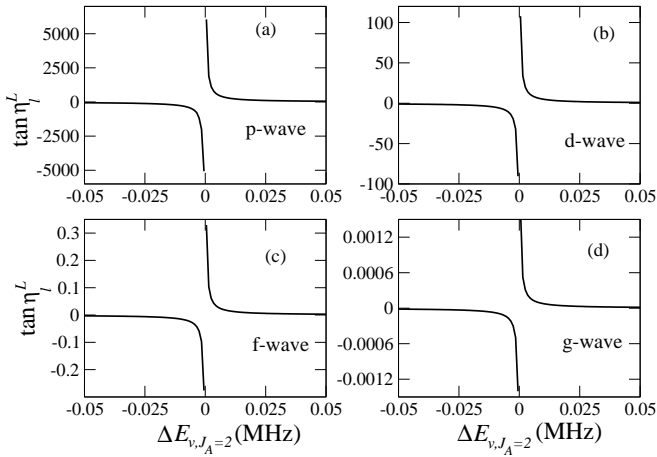}
\caption{$\tan \eta_{\ell}^{L}$ ($\eta_{\ell}^{L}$ is the light induced phase-shift) is plotted as a function of detunning $\Delta E_{v,J_A=2}$ (MHz) when L$_A$ is tuned near $J_A=2$. The total shift $E_{J_A = 2}^{shift}$ at 40 kW/cm$^2$ is -0.91 GHz. The other parameters are $I_{A}$ = 40 kW/cm$^2$ and $E$ = 10 $\mu$K}
\end{figure}

\begin{figure}
\includegraphics[width=2.50in]{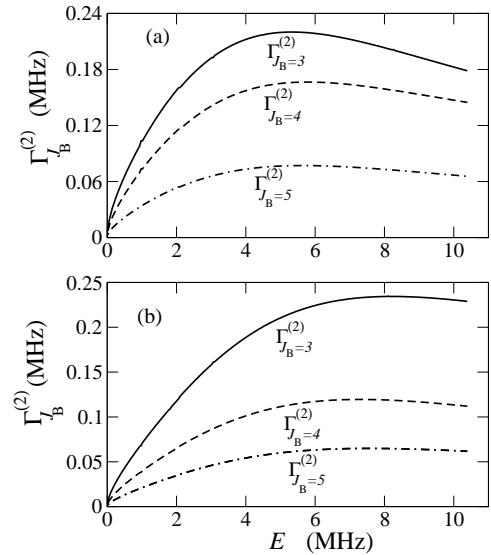}
\caption{The Two-color total stimulated line width $\Gamma^{(2)}_{J_B}$ (in unit of MHz) for different $J_B$ (as indicated in the plots) is plotted as a function of collisional energy $E$ (in unit of MHz) when L$_A$ is tuned near $J_A$ = 1 for $\delta_A$ = -1.25 GHz (a),  $\delta_A$ = -1.48 GHz (b) with $I_A$ = 40 kW/cm$^2$ and $I_B$ = 1 W/cm$^2$.}
\end{figure}

For comparison, we also calculate one-color partial stimulated line widths $\Gamma_{J\ell}^{(0)}$ for $J > 2$ from the expression $\Gamma_{J\ell}^{(0)} = 2\pi|f_{J M; \ell m_{\ell}}|^2 \label{partialgama0}$. The one-color total stimulated line width is $\Gamma_J^{(0)} = \sum_{\ell, m_{\ell}, M} \Gamma_{J\ell}^{(0)} $. At 10 $\mu$K energy and at laser intensity 1 W/cm$^2$, the one-color partial stimulated line widths $\Gamma^{(0)}_{J=3,\ell=1}$ = 15.46 Hz, $\Gamma^{(0)}_{J=4,\ell=2}$ $\simeq$ 0, $\Gamma^{(0)}_{J=5,\ell=3}$ $\simeq$ 0. A comparison between one- and two-color partial stimulated line widths has been made in Table II for $J_B > 2$ at collisional energy 10 $\mu$K. The two-color total line widths $\Gamma^{(2)}_{J_B=3}$, $\Gamma^{(2)}_{J_B=4}$, $\Gamma^{(2)}_{J_B=5}$ are  0.03537 MHz, 0.0200 MHz and 0.0103 MHz, respectively when $\delta_{A}$ = -1.25 GHz  and they are 0.02229 MHz, 0.0122 MHz and 0.0061 MHz, respectively for $\delta_{A}$ = -1.48 GHz. The corresponding one-color weak-coupling partial as well as total stimulated line widths $\Gamma^{(0)}_{J\ell}$ and $\Gamma^{(0)}_{J}$ for the same rotational states with laser intensity of 1 W/cm$^2$ are vanishingly small while the two-color partial $\Gamma^{(2)}_{J_B\ell}$ and total $\Gamma^{(2)}_{J_B}$ exceed $\Gamma^{(0)}_{J\ell}$ and $\Gamma^{(0)}_{J}$ by several orders of magnitude. We find the energy shift $|E^{shift}_{J_A=1}|$ is 0.79 GHz which exceeds the spontaneous line width $\gamma$ (say 2 MHz for model calculation) by two orders of magnitudes.
\begin{table}
\caption{Tabulated are the $\tan \eta_{\ell}^{L}$ when the laser $L_A$ is tuned near $J_A = 2$ at $E$ = 10 $\mu$K for three values of  $\delta_{A}$. The parameters are  $I_A = 40 $ kW/cm$^2$, $E_{vJ_A = 2} = -2.138$ GHz and $E_{J_A = 2}^{shift} = - 0.91$ GHz. In the field-free case, $\tan \eta_{\ell=1}^{0}$ =  -1.57$\times 10^{-4}$, $\tan \eta_{\ell=2}^{0}$ = 1.20$\times 10^{-6}$, $\tan \eta_{\ell=3}^{0} \simeq$ 0 and $\tan \eta_{\ell=4}^{0} \simeq$ 0}
\begin{tabular}{c  c  c  c  c  c  c}
\hline 
\multicolumn{1}{c}{} & \multicolumn{1}{c}{\vline} & \multicolumn{1}{c} {$\delta_{A}$ = -2.95 GHz} &  \multicolumn{1}{c}{\vline} & \multicolumn{1}{c} {$\delta_{A}$ = -3.049 GHz} &  \multicolumn{1}{c}{\vline}  & \multicolumn{1}{c} {$\delta_{A}$ = -3.17 GHz} \\
\hline
\multicolumn{1}{c}{$\ell$} &  \multicolumn{1}{c}{\vline}  & \multicolumn{1}{c} {$\tan \eta_{\ell}^{L} \times 10^{4}$} &  \multicolumn{1}{c}{\vline} & \multicolumn{1}{c} {$\tan \eta_{\ell}^{L} \times 10^{4}$} &  \multicolumn{1}{c}{\vline}  & \multicolumn{1}{c} {$\tan \eta_{\ell}^{L} \times 10^{4}$} \\
\hline
1  & \vline & 263.00 & \vline & 36900.00 & \vline & -229.00  \\
2  & \vline & 4.93 & \vline & 659.00 & \vline & -4.09 \\
3  & \vline & 0.01 & \vline & 2.01 & \vline & -0.01 \\
4  & \vline & 0.00 & \vline & 0.01 & \vline & 0.00  \\
\hline
  \end{tabular}
\label{tb2}
\end{table}

In order to trace the origin of increment of $\Gamma^{(2)}_{J_B\ell}$ we plot perturbed $\psi_{E \ell m_{\ell}}$ for $\ell \neq 0$ when laser L$_A$ is tuned near $J_A = 1$ and the corresponding field-free regular functions $\psi^{0,reg}_{E\ell}$ in Fig.3. It is clear from this figure that the amplitudes of $\psi_{E \ell m_{\ell}}$ are greatly enhanced by several orders of magnitude than that of $\psi^{0,reg}_{E\ell}$. Next, we calculate $\tan \eta_{\ell}^{L}$ by using Eq. (\ref{newphase}) when laser L$_A$ is tuned near $J_A=2$. These are given in Table III for $\delta_{A}$ = -2.95 GHz, -3.17 GHz and -3.049 GHz. The first two $\delta_{A}$ values correspond to off-resonant and the last one to resonant condition. The variation of $\tan \eta_{\ell}^{L}$ with $\Delta E_{v,J_A = 2}$ is plotted in Fig.4 which exhibits resonance for higher partial waves induced by strong-coupling PA. The enhancement of the partial ($\ell\neq 0$) wave amplitude is due to the term $\sum_{\ell' m_{\ell'} M } A_{ J M; \ell' m_{\ell'}}(E)$ of Eq. (\ref{nscf1}). In figure 5, the two-color total stimulated line width $\Gamma^{(2)}_{J_B}$ is plotted as a function of collisional energy $E$ for two off-resonant $\delta_A$ values when L$_A$ is tuned near $J_A$ = 1. The magnitude of $\Gamma^{(2)}_{J_B}$ for higher rotational states ($J_B= 4, 5$) is less than that of $J_B=3$. This is due to the fact that the lowest possible partial wave contribution to the excitation of rotational states $J_B=4$ and $J_B = 5$ are $d$ and $f$, respectively while $J_B=3$ state can be populated from $p$ wave which has rotational barrier lower than that of d- and f-wave. The two-color photoassociation rate $K^{(2)}_{PA}$ as defined in Eq.(\ref{Photorate}) has been plotted as a function of $\delta_B$ (Fig.6a) and $\Delta E_{v,J_B}$ (Fig.6b). The spectra in Fig.6b are red-shifted due to the presence of the term $\Gamma^{(2)}_{J_B}$ in Eq. (\ref{Photorate}). From the selection rule, it is obvious that $J_B = 3, 4, 5 $ rotational states can not be populated by a PA transition from s-wave scattering state. But the appearance of the $J_B = 3, 4, 5$ lines in PA spectra is an indication of the significant modification of the partial scattering wavefunctions by intense light field.  
\begin{figure}
\includegraphics[width=2.60in]{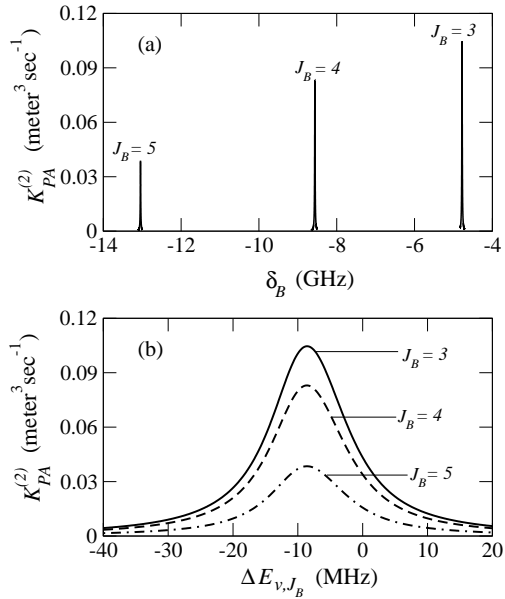}
\caption{The upper panel (a) shows two-color photoassociation rate $K_{PA}^{(2)}$ (in unit of meter$^{3}$ sec$^{-1}$) as a function of atom-field detunning  $\delta_B$ (in unit of GHz) for three higher rotational levels $J_B$ (as indicated in the plots) for $\delta_A$ = -1.25 GHz  when the laser L$_A$ is tuned near $J_A=1$. The lower panel (b) shows the same but as a function of detunning $\Delta E_{v,J_B} = \hbar\delta_B-E_{v,J_B}-E_{J_B}^{shift}$ (in unit of MHz) from PA resonance. The other parameters for both the panels are $I_{A}$ = 40 kW/cm$^2$, $I_{B}$ = 1 W/cm$^2$ and $T$ = 100 $\mu$K.}  
\end{figure}

\section{CONCLUSION}
In the present paper we have developed a two-color PA scheme for the excitations of higher ($ J_B > 2$) rotational levels which are generally suppressed in the Wigner threshold law regime. We have calculated two-color stimulated line width (for $J_B > 2$) by fixing strong laser either near $J_A = 1$ or $ J_A = 2$ state and tuning another weak laser to higher rotational ($J_B =3, 4, 5$) states. Then we have compared these with one-color line widths. The enhancement of stimulated line width is a result of strong-coupling photoassociative dipole interaction which in turn modifies the continnum states. This proposed method may be important for coherent control of rotational excitations and manipulation of optical Feshbach resonance of higher partial waves.

\appendix
\section{}

The mathematical treatment given here is closely related to our earlier work \cite{debPRL10409}. Treating the laser field classically, the effective interaction Hamiltonian under rotating wave approximation the two state basis can be expressed as
\bea
H_{eff}^{int} = \exp(-i\delta t) \Omega_{eg}(r) | e \rangle \langle g | + \rm{H.c.}
\label{heff} \eea
Form time-independent Schr\"{o}dinger equation $H\Psi_{E} = E\Psi_{E}$, we obtain two coupled equations 
\bea
[ - \frac{\hbar^2}{2\mu}\nabla_r^2 + V_{\rm g}(r) - E ] \Phi_g({\mathbf r}) = - \Omega_{ge}(r) \Phi_e({\mathbf r})
\label{3dg}
\eea
\bea
[ - \frac{\hbar^2}{2\mu}\nabla_r^2 + V_{\rm ex}(r) - E - \hbar \delta ] \Phi_e({\mathbf r}) = - \Omega_{eg}(r) \Phi_g({\mathbf r})
\label{3de}
\eea
Here $V_{\rm g}$ is assumed to include hyperfine interaction of the chosen channel. Substituting  Eqs. (\ref{grstate}) and (\ref{exstate}) into the Sch\"{o}dinger equations (\ref{3dg}) and (\ref{3de}) we get two coupled equations
\bea
\left [ -\frac{\hbar^2}{2\mu}\frac{d^2}{d r^2} + B_J(r) + V_{\rm ex}(r) - \hbar \delta - E - i \hbar \frac{\gamma}{2} \right ]  \phi_{v J} 
  &=& \nonumber \\ - \sum_{\ell m_{\ell}}\Lambda_{JM;\ell m_{\ell}} \tilde{\psi}_{E \ell m_{\ell}}
\label{coup1}\eea
\bea
\left [ - \frac{\hbar^2}{2\mu}\frac{d^2}{d r^2} + B_{\ell}(r) + V_{\rm g}(r) - E \right]\tilde{\psi}_{E \ell m_{\ell}} 
  &=& \nonumber \\ -\sum_{M}  \Lambda_{\ell m_{\ell}; J M} \phi_{v J}
\label{coup2}
\eea
where  $B_J(r) = \hbar^2/(2\mu r^2) [J(J+1) - \Omega^2]$ is the rotational term of excited molecular bound state in the absence of nuclear spin, $B_\ell(r) =  \hbar^2/(2\mu r^2) \ell(\ell +1)$ is the centrifugal term in collision of two ground state (S) atoms, $\tilde{\psi}_{E \ell
m_{\ell}}(r) =  \int_{E'} \beta_{E'} \psi_{E' \ell m_{\ell}}( r )d
E'$. The above two equations are solved by the green's function method by setting $\Lambda_{J M;\ell m_{\ell}} = \Lambda_{\ell m_{\ell}; J M} = 0$. The single channel scattering  equation becomes
\be
\left [ - \frac{\hbar^2}{2\mu}\frac{d^2}{d r^2} + B_{\ell}(r) + V_{\rm g}(r) -E\right]\psi_{E\ell}^{0} = 0.
\label{homo2}\ee
Let $\psi_{E \ell}^{0 , reg}(r)$ and $\psi_{E \ell}^{0 , irr}(r)$ represent the regular and irregular solutions of the above equation. 
The appropriate Green's function for the scattering wave function can be written as
\bea
{\cal K}_{\ell} (r,r') = -\pi \psi_{E \ell}^{0 ,reg}(r)\psi_{E \ell}^{0 ,irr}(r')  \hspace{0.5cm}  (r'>r) \eea
\bea
{\cal K}_{\ell} (r,r') = -\pi \psi_{E \ell}^{0 , reg}(r')\psi_{E \ell}^{0 , irr}(r) \hspace{0.5cm}  (r'<r)
\eea
The regular function, $\psi_{E \ell}^{0 , reg}(r)$ vanishes at r=0 and the irregular solution $\psi_{E \ell}^{0 , irr}(r)$ is defined by boundary only at $r\rightarrow\infty$. The energy normalised asymptotic form of both regular and irregular wave function is 
\bea
\psi_{E\ell}^{0,reg} = \sqrt{\frac{2\mu}{\pi \hbar^2 k}} \sin(kr-\frac{\ell \pi}{2} +\eta_{\ell}^{0}), \hspace{0.5cm} r \rightarrow \infty
\eea
\bea
\psi_{E\ell}^{0,irr} = \sqrt{\frac{2\mu}{\pi \hbar^2 k}} \cos(kr-\frac{\ell \pi}{2} +\eta_{\ell}^{0}), \hspace{0.5cm} r \rightarrow \infty
\eea where $\eta_{\ell}^{0}$ is the phase-shift of $\ell$-th partial wave in the absence of PA coupling. The homogeneous part of (\ref{coup1}) with $\gamma$ = 0 is
\be
\left [ -\frac{\hbar^2}{2\mu}\frac{d^2}{d r^2} + B_J(r) + V_{\rm ex}(r) \right ] \phi_{vJ}^{0} =(\hbar\delta+E)\phi_{vJ}^{0} = E_{vJ} \phi_{vJ}^{0}
\label{homo1}\ee  The Green function corresponding to these rovibrational states $\phi_{vJ}^{0}$ can be written as \be G_{v}(r,r')= - \frac{1}{\hbar \delta + E - E_{vJ} + i\hbar \gamma/2 }\phi_{vJ}^{0}(r)\phi_{vJ}^{0}(r')
\ee Using this Green's function, we can write down the solution of equation (\ref{coup1}) in the form
\bea
\phi_{vJ} (r) &=& -\sum_{\ell m_{\ell}} \int dr'\Lambda_{JM; \ell
m_{\ell}}(r')G_{v}(r,r')\tilde{\psi}_{\ell m_{\ell}}(r') \nonumber \\ &=& \int_{E'} \beta_{E'}\sum_{\ell m_{\ell}} A_{JM; \ell
m_{\ell}} \phi_{vJ}^{0}(r) dE'
\label{coupb}\eea where \bea A_{JM; \ell m_{\ell}} &=& \sum_{\ell m_{\ell}} \int dr'\Lambda_{JM; \ell
m_{\ell}}(r')\phi_{vJ}^{0}(r') \psi_{E \ell m_{\ell}}(r')\nonumber \\ &\times& \frac{1}{\hbar \delta + E - E_{vJ} + i\hbar \gamma/2 }\label{VA}\eea Substituting equation (\ref{coupb}) into equation (\ref{coup2}) we obtain \bea \left [ \frac{\hbar^2}{2\mu}\frac{d^2}{d r^2} - B_{\ell}(r) - V_{g}(r)+E \right ]\psi_{E \ell m_{\ell}}(r) &=& \nonumber \\  \sum_{\ell^{'} m_{\ell^{'}} M} A_{JM; \ell^{'} m_{\ell^{'}}}\Lambda_{\ell
m_{\ell}: J M}(r)\phi_{vJ}^{0}(r)
\label{coupg}\eea The scattering solution can now be expressed as
\bea
\psi_{E \ell m_{\ell}}( r ) &=& \psi_{E\ell}^{0,reg} + \sum_{\ell' m_{\ell'} M } A_{ J M; \ell' m_{\ell'}}(E) \nonumber \\&\times& \int {\cal K_{\ell}}(r,r')\Lambda_{ \ell m_{\ell}; J
M}(r')\phi_{vJ}^{0}(r')dr' \label{nscf}\eea 
On substitution of equation (\ref{nscf}) into (\ref{VA}) and after some algebra, we obtain
\begin{widetext}
\be A_{J,M; \ell, m_{\ell}} =  \frac{1}{\hbar\delta + E - E_{vJ} + i \hbar \gamma/2 - E_J^{{\mathrm shift}}}  \times \left [ f_{J M : \ell m_{\ell}} + E_{J \ell}^{shift} \sum_{\ell^{'} \ne \ell,m_{\ell^{'}} M'} A_{JM':\ell^{'}m_{\ell^{'}}}\right ] \label{AJ}\ee
\end{widetext} Let $D = \hbar\delta + E - E_{vJ}-E_{J \ell}^{shift}+i\hbar\gamma/2$. Now, adding a term $D^{-1} E_{J \ell}^{shift}  A_{J,M; \ell, m_{\ell}} $ on both side of equation (\ref{AJ}), we can express $A_{J,M; \ell, m_{\ell}}$ in terms a quantity $\tilde{A}_J = \sum_{\ell m_{\ell} M }A_{J,M; \ell, m_{\ell}}$ as well as other parameters. On summing over all possible $\ell, m_{\ell}, M$ we can evaluate $\tilde{A}_J$. Having done all these algebra, we can explicitly express \bea A_{J,M; \ell, m_{\ell}} &=& \frac{\left [ f_{J, M; \ell m_{\ell}} +  E_{J \ell}^{shift}  \tilde{A}_J \right ]}{\hbar\delta+E-E_{vJ} + i \hbar \gamma/2} \eea and \be \tilde{A}_J = \sum_{\ell, m_{\ell}, M} \frac{
f_{J M; \ell m_{\ell}}}{\hbar\delta+E-E_{vJ} + i \hbar \gamma/2 -
E_{J}^{shift}} \ee

\end{document}